\documentclass[aps, pre, twocolumn,a4paper,showkeys,showpacs]{revtex4-1}
\usepackage[dvips]{graphicx}
\usepackage{amsmath}
\usepackage{bm}
\usepackage{color}
\usepackage{natbib}
\usepackage{comment}
\usepackage[utf8]{inputenc}
\usepackage{MnSymbol}

\begin{document}
\title{Collective Cell Movement in Cell-Scale Tension Gradient on Tissue Interface}
\author{Katsuyoshi Matsushita$^1$, Hidenori Hashimura$^2$, Hidekazu Kuwayama$^3$, Koichi Fujimoto$^1$}

\affiliation{$^1$Department of Biological Sciences, Osaka University, Toyonaka, 560-0043, Osaka, Japan\\
$^2$Graduate School of Arts and Sciences, University of Tokyo, Komaba, 153-8902, Tokyo, Japan\\
$^3$Faculty of Life and Environmental Sciences, University of Tsukuba, Tsukuba, 305-8572, Ibaraki, Japan}

\begin{abstract}
In this paper, we examine the emergence of cell flow induced by a tension gradient on a tissue interface as in the case of the Marangoni flow on liquid interface. We consider the molecule density polarity of the heterophilic adhesion between tissues as the origin of the tension gradient. By applying the cellular Potts model, we demonstrate that polarization in concentration (i.e., intracellular localization) of heterophilic adhesion molecules can induce a cell flow similar to the Marangoni flow. In contrast to the ordinary Marangoni flow, this flow is oriented in the opposite direction to that of the tension gradient. The optimal range of adhesion strength is also identified for the existence of this flow.
\end{abstract}

\maketitle

\section{Introduction}

The collective migration of Eukaryotic cells attracts the attention of physicists as one of non-equilibrium collective phenomena \cite{Hakim:2017}.  In particular, the driving mechanism of the collective migration have been investigated with focus on their various microscopic cell-scale factors, including the chemotaxis \cite{Weijer:2009}, cytoskeleton contraction\cite{Rauzi:2008}, contact inhibition of locomotion \cite{Carmona-Fontaine:2008}, cell-substrate adhesion \cite{Pascalis:2017}, and cell--cell adhesion \cite{Takeichi:2014}.  The roles of these factors are well explained through theoretical reproduction of the collective migration in a macroscopic scale \cite{Rappel:1999, Szabo:2006, Lober:2015, Sato:2015a, Camley:2016, Najem:2016, Sato:2017, Campo:2019, Oelz:2019, Hiraiwa:2020, Alert:2020, Okuda:2021}.
The cell--cell adhesion has been secondarily considered as a stabilizer of cell--cell contacts, but it is not direct driving force of the collective cell migration \cite{Lee:2011a, Kabla:2012}.
Here, we theoretically propose a possibility of the cell-cell adhesion as the driving force from the physics point of view.

\begin{figure}[t]
    \begin{center}
        \includegraphics[width=1\linewidth]{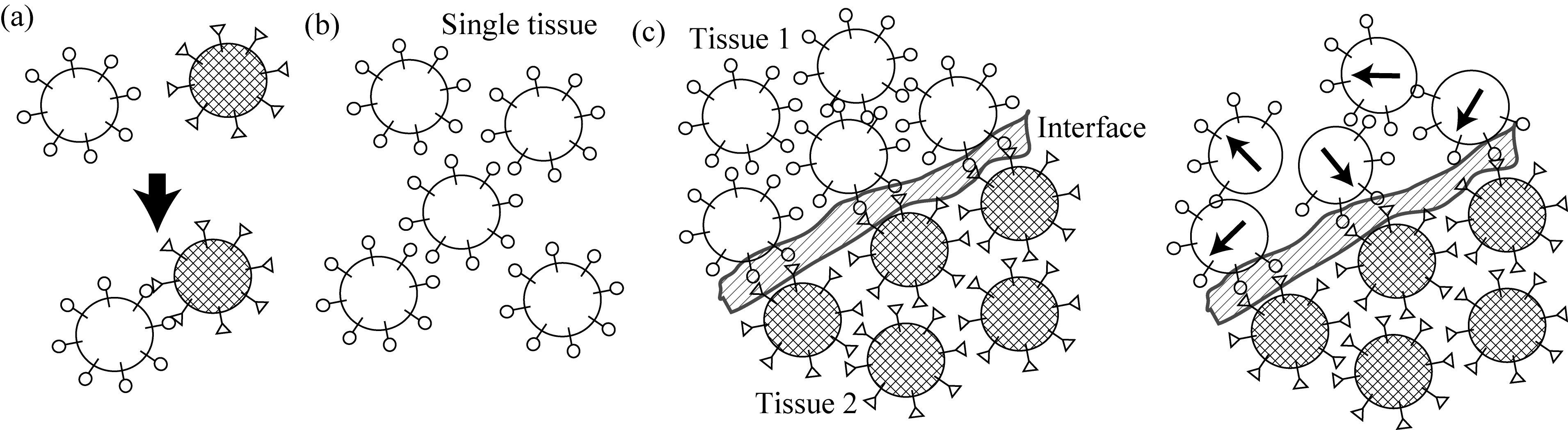}
        \caption{The schematic view of heterophilic cell--cell adhesion. White and shaded circles represent cells with different heterophilic adhesion molecules on their membranes. Circular and triangle symbols with a line represent two different heterophilic adhesion molecules, which stabilize a cell--cell contact. (a) The stabilization of cells is due to heterophilic adhesion. (b) The case of tissue with only one of heterophilic molecules. (c) The case of an interface between two tissues, which have either of heterophilic adhesion molecules respectively.
        (d) The case of tissues with surface tension gradient on each cell surface in one of tissues.}
        \label{fig:adhesion}
    \end{center}
\end{figure}

Cell--cell adhesion regulates various cellular processes and promotes tissue organization by stabilizing mechanical contacts between cells \cite{Takeichi:2014}.
This adhesion results from molecular adhesion binding between the surface membrane of two cells
One typical  binding appears between different adhesion molecules, as shown in Fig.~\ref{fig:adhesion}(a), called heterophilic adhesion.
As the particular characteristics of this adhesion, when only either of two heterophilic adhesion molecules exists in a  cell population (tissue) shown in Fig.~\ref{fig:adhesion}(b), the adhesion does not affect the tissue.
In contrast, when each and every one of heterophilic adhesion molecules separately exists only on one of two tissues, respectively, as shown in Fig.~\ref{fig:adhesion}(c), the heterophilic adhesion can regulate the tension at the interface of the tissues.
Therefore, this adhesion is expected to be effective in the regulation of tissue interfaces with avoiding side effects in each tissue.
In particular, when the molecules induce the gradient of the interface tension by polarization in their molecule concentrations on each cell surface \cite{Sesaki:1996,Coates:2001} as shown in Fig.~\ref{fig:adhesion}(d), we expect the interface regulation using the Marangoni effect of each cell on the interface. 
In this case, the cell-scale Marangoni effect becomes the physical mechanism of collective cell motions for the heterophilic cell--cell adhesion.

\begin{figure}[t]
    \begin{center}
        \includegraphics[width=0.75\linewidth]{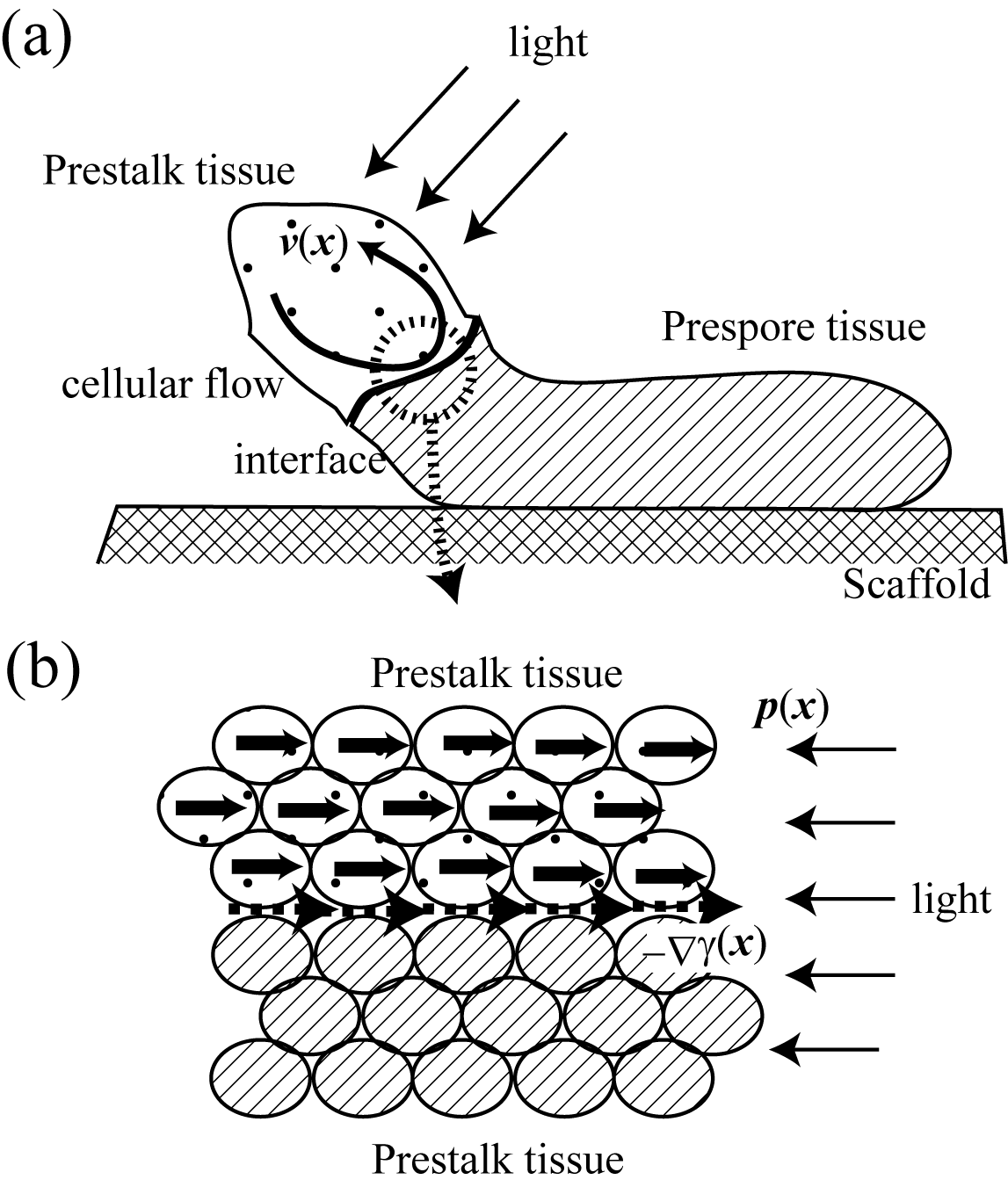}
        \caption{The hypothetical mechanism of collective cell flow during phototaxis of the slug. (a) The schematic structure of dicty slug. The dotted and shaded  regions represent the prestalk (leading) and prespore (following) tissues, respectively. The cross-hatched region represents the scaffold, which is where the dicty are manifested. In the leading tissue, cells convectively flow with the slug’s phototaxis. (b) A magnified view of the region surrounding the interface between the two tissues. The dotted and dashed circles represent the prestalk and prespore cells, respectively. The prestalk cells are assumed to be polarized in concentration of a heterophilic adhesion molecule responding to light. Here, the solid arrows in the prestalk cells represent the polarization in concentration,  $\bm p(x)$ , and the dotted arrows on the tissue interface represent the tension gradient, $-\nabla \gamma(\bm x)$.}
        \label{fig:phototaxis}
    \end{center}
\end{figure}

This mechanism of collective cell motion is expected in the slug stage of {\it Dictyostelium discoideum} (dicty) \cite{Bonner:2009}.
At this stage, dicty cells are differentiated into two types—prestalk and prespore. In particular, the prestalk cells form the tissue in the leading region of the slug. They sometimes convectively flow during the movement of the slug in response to light, as illustrated in Fig.~\ref{fig:phototaxis}(a) \cite{Kimura:2000,Hashimura:2019b}. 
This flow is speculated to regulate the slug’s phototaxis by inducing the exertion of a torque on the leading region of the slug. For a long time, the flow was hypothesized to be an effect of chemotaxis. However, recent observations have revealed that chemotaxis is inert at this stage \cite{Hashimura:2019a}. 
Further, a chemotaxis-deficient mutant of dicty, KI5  \cite{Kuwayama:1993, kuwayama:1995,kuwayama:2013} has exhibits normal slug movement \cite{Kida:2019}. 
 Instead of chemotaxis, we contend that the interface tension gradient between prestalk and prespore tissues acts as the driving force of this flow. In particular, the tissue interface tension, $\gamma(\bm x)$, induces a flow,  $\bm v(\bm x)$, in the interface  \cite{Levich:1969,Brochard:1989,Getling:1998,Squires:2005}:
\begin{align}
    \bm v(\bm x) \propto \nabla \gamma(\bm x). \label{eq:Marangoni_flow}
\end{align}
Here, $\bm x$ denotes a position on the interface.
 In this paper, we hypothesize that this gradient is induced by the spontaneous polarization in heterophilic adhesion in response to light in prestalk cells, as depicted in Fig.~\ref{fig:phototaxis}(b). In this case, the polarization vector, $\bm p(\bm x)$ induces the tension. Thus, we have:
\begin{align}
    - \nabla \gamma(\bm x) \propto \bm p(\bm x). \label{eq:polarization-tension}
\end{align}

A hypothetical scenario of this gradient is schematically illustrated in Fig.~\ref{fig:phototaxis}(b).
The concentration of heterophilic adhesion molecules in the prestalk tissue is assumed to be polarized in the direction of a light source.
This polarization, $\bm p(\bm x)$, induces a tension gradient, $-\nabla \gamma(\bm x)$, along the interface. 
This gradient drives the collective cell flow near the interface as a relative motion between two tissues and, thereby, a cell flow in the prestalk tissue. With respect to the common origin of the tension gradient, this effect is similar to the Marangoni flow on liquid interfaces with a tension gradient  \cite{Levich:1969,Getling:1998}. 
Therefore, this cell flow may be called “cell Marangoni flow”. Unlike the Marangoni flow, the tension gradient,  $-\nabla \gamma(\bm x)$, is manifested on the cell scale \cite{Coates:2001} because the polarization is expected only in individual cells. Further, cells flow via their shape deformations, which are absent in a liquid. Therefore, the polarization in heterophilic adhesion does not simply result in a Marangoni flow. To address these concerns in our scenario, the theoretical confirmation of the flow based on the scenario is at least necessary. 

In this paper, we undertake a theoretical examination of the “cell Marangoni flow” based on the aforementioned scenario. To this end, we consider the two-dimensional cellular Potts model \cite{Graner:1992,Graner:1993,Glazier:1993} consisting of two tissues, which correspond to the two different types of molecules participating in heterophilic cell-cell adhesion. 
We demonstrate that the polarization in heterophilic cell-cell adhesion in one tissue induces the relative motion between two tissues. In contrast to the Marangoni flow
 \cite{Getling:1998}, the direction of this flow is aligned to that of the tension gradient and against the direction of the Marangoni flow. This is expected as low and high tensions promote cell-shape extension and shrinkage \cite{Matsushita:2017}, respectively, during cell movement. Further, we investigate this flow as a function of adhesion strength and schematically clarify the steady states. Based on this clarification, we determine the emergence condition of the flow in heterophilic adhesion.

\section{Model}

The cellular Potts model is a variant of the Potts model and is widely used to express cellular dynamics \cite{Scianna:2013, Hirashima:2017}.
As the effects of cell-cell adhesion is easily expressible using this model compared to others \cite{Anderson:2007},
it is particularly useful for our examination of heterophilic cell-cell adhesion.
The state space of this model expresses the space of cell configurations, each of which is represented by a Potts state in the lattice.
A Potts state, $m(\bm r)$ is defined corresponding to each lattice site, $\bm r$ and it takes a value in $\{0, 1, \dots, N\}$.  The value of  $m(\bm r)$ represents the index of the cell that occupies $\bm r$, when $m(\bm r)$ $\not = 0$. In contrast, $m(\bm r)$ = 0 denotes that the site, $\bm r$, is empty. The largest index, $N$, is equal to the number of cells. Based on this interpretation, the Potts state,  $\{m(\bm r)\}$, expresses cells as the domains of corresponding Potts states and, thereby, cell configurations.

In a Potts state, cells are classified into types with different features \cite{Graner:1992}.  
In the model proposed in this paper, two categories of cells are introduced to represent the two tissues corresponding to the two different heterophilic adhesion molecules binding with each other. Each model cell is of either of these two types and can participate in heterophilic adhesion only with cells of the other type. The type of the $m$th cell is denoted by $T(m)$.
$T(m) = 1$ corresponds to a polarized concentration of heterophilic adhesion molecules and $T(m) = 2$ corresponds to an isotropic concentration. In the dicty slug depicted in Fig.~\ref{fig:phototaxis}(b), the cells with $T(m) = 1$ correspond to the prestalk tissue and those with $T(m) = 2$ correspond to the prespore tissue.

For convenience during the construction of this model, the type function, $T(m)$, is further extended to Potts states that do not correspond to real cells. For an empty space with  $m$ = 0, we set $T(0)$ = 0.
 In addition, we introduce a fixed scaffold on which the cells live. The motivation behind this is to inhibit artificial translational motions of the whole system, which give rise to systematic noise during the analysis of collective cell motions \cite{Matsushita:2020}. 
To this end, we consider fixed cells with $T(m)$ = 3  and define some cells of this type to constitute the scaffold. In this case, we assume cells with $T(m)$ = 2  to be highly amenable to adhesive contact with the scaffold and cells with $T(m)$ = 1 to be incapable of such contact. 
In the dicty slug, these configurations correspond to the situation in which only the tissue consisting of prespore cells preferably makes contact with the scaffold due to its low surface tension. Henceforth, the extended type is denoted by the italicized capital letter,  $T$. In summary, $T$ takes the value 0 corresponding to empty spaces, 1 corresponding to tissues with $T(m)$ = 1, 2 corresponding to tissues with $T(m)$ = 2, and 3 corresponding to the scaffold.

The Potts state $\{m(\bm r)\}$ is repeatedly updated using a Markov Chain Monte Carlo simulation and is regarded as a snapshot in a time series of moving cells. In this case, the occurrence probability of  $\{m(\bm r)\}$ is given by the Boltzmann factor, $P(\{m(\bm r)\})$ = $\exp\{-\beta {\cal H}(\{m(\bm r)\})\}$ / $\sum_{\{m(\bm r)\}}\exp\{-\beta {\cal H}(m(\bm r))\}$. 
Here, $\beta$ denotes an inverse temperature representing fluctuations in cell shapes and $\cal H$ denotes the free energy, which is defined to be the summation of the following four terms:
\begin{align}
    {\cal H}(m(\bm r)) = {\cal H}_{\rm Ten}  + {\cal H}_{\rm Hom} + {\cal H}_{\rm Het} + {\cal H}_{\rm Are}. \label{eq:Free_enegy}
\end{align}

The first term in the right hand side (RHS) of Eq.~\eqref{eq:Free_enegy} can be further decomposed into two terms:
\begin{align}
    {\cal H}_{\rm Ten} = {\cal H}_{\rm E} + \sum_{T=1}^2 {\cal H}_{{\rm S}}^T. \label{eq:interaction_medium}
\end{align} 
The first term in RHS of Eq.~\eqref{eq:interaction_medium} represents the surface tension between cells and
empty spaces. ${\cal H}_{\rm E}$ is defined as follows:
\begin{align}
    {\cal H}_{E} = \Gamma_{\rm E} \sum_{T=1}^2\sum_{\bm r \bm r'} \left(\delta_{T(m(\bm r))T}\delta_{T(m(\bm r'))0} + \right. \nonumber \\
    +  \left.\delta_{T(m(\bm r'))T}\delta_{T(m(\bm r))0} \right).
\end{align} 
Here, $\delta_{ab}$  denotes the Kronecker $\delta$. The summation with respect to the pair, $\bm r$ and  $\bm r'$, is taken over the nearest and next nearest sites. This summation rule is also applied to all equations that appear hereafter. The surface tension,  $\Gamma_{\rm E}$, is assumed to be identical for $T(m)$ = 1 and $T(m)$ = 2, for simplicity.
The second term in the RHS of Eq.~\eqref{eq:interaction_medium} represents the surface tension between cells and scaffolds.
Here, ${\cal H}_{{\rm S}}^T$ is given by
\begin{align}
    {\cal H}_{{\rm S} }^T =  \Gamma_{{\rm S}}^T \sum_{\bm r \bm r'} \left(\delta_{T(m(\bm r))T}\delta_{T(m(\bm r'))3} + \right. \nonumber \\
    +  \left.\delta_{T(m(\bm r'))T}\delta_{T(m(\bm r))3} \right).
\end{align} 
The surface tension with the scaffold, $\Gamma_{{\rm S}}^T$, depends on the type of the cell, $T$. 
$\Gamma_S^2$ is assumed to be significantly lower than $\Gamma_S^1$ to capture the relative ease with which cells with $T$ = 2 establish contact with the scaffold compared to those with $T$ = 1.

The second term in the RHS of Eq.~\eqref{eq:Free_enegy} is given by
\begin{align}
    {\cal H}_{\rm Hom} = \sum_{T=1}^{2} \Gamma^{T} \sum_{\bm r \bm r'}\delta_{T(m(\bm r))T} \delta_{T(m(\bm r'))T} \eta_{m(\bm r) m(\bm r')} \nonumber\\
    + \frac{\Gamma_{\rm I}}{2} \sum_{T\not=T}\sum_{\bm r \bm r'}\delta_{T(m(\bm r))T} \delta_{T(m(\bm r'))T'}. \label{eq:H_hom}
\end{align}
The first and second terms in the RHS of Eq.~\eqref{eq:H_hom} represent the interface tension between cells of identical types and those of different types, respectively. Further, 
$\eta_{ab}$ = 1 $-$ $\delta_{ab}$. The interface tensions, $\Gamma^T$ ($T$ = 1 or 2) and $\Gamma_{\rm I}$ , correspond to those of homophilic adhesion between cells of identical and different types, respectively. Homophilic adhesion stabilizes the tissues.

The third term in the RHS of Eq.~\eqref{eq:Free_enegy} is given by
\begin{align}
    {\cal H}_{\rm Het} &= - \sum_{\bm r \bm r'} \Gamma_{\rm Het}(\bm r, \bm r') \nonumber \\
   &\times (\delta_{T(m(\bm r))1}\delta_{T(m(\bm r'))2}+\delta_{T(m(\bm r))2}\delta_{T(m(\bm r'))1}),
\end{align}
and it establishes heterophilic adhesion on the tissue interface to be the driving force of the flow. Here, $\Gamma_{\rm Het}(\bm r, \bm r')$  denotes the reduction in tissue interface tension induced by heterophilic adhesion between two cells occupying the sites, $\bm r$ and $\bm r'$.

$\Gamma_{\rm Het}(\bm r, \bm r')$ denotes the polarization in adhesion \cite{Zajac:2002, Vroomans:2015, Matsushita:2017, Matsushita:2018}. 
In this expression, $\Gamma_{\rm Het}(\bm r, \bm r')$ is assumed to depend on the concentrations of the adhesion molecule, $\rho_{m({\bm r})}({\bm r})$ and $\rho_{m({\bm r'})}({\bm r'})$, on a microscopic level \cite{Matsushita:2017}. Here, $\rho_{m}({\bm r})$ denotes the concentration of the adhesion molecule at $\bm r$ in the $m$th cell. In the leading order terms of these concentrations, we assume the surface tension to be given by:
\begin{eqnarray}
 \Gamma_{\rm Het}(\bm r, \bm r') = \varepsilon \rho_{m(\bm r)}(\bm r)\rho_{m(\bm r')}(\bm r'). \label{eq:g-het}
\end{eqnarray}
This equation qualitatively realizes that the surface tension is reduced by 
the heterophilic adhesion molecule binding between $\bm r$ and $\bm r'$.

To further introduce the concept of polarization in heterophilic adhesion, we consider the multipole expansion of $\rho(\bm r)$ \cite{Arfken:2012, Marchetti:2013, Matsushita:2017}:
\begin{eqnarray}
    \rho_m(\bm r) = \rho_m^{\rm M} + \rho_m^{\rm D}(\bm e_m(\bm r)\cdot \bm p_m)+ \dots.
\end{eqnarray}
Then, we utilize the terms up to the order of the dipole part of $\rho_m^{\rm D}$ corresponding to  $T(m)=1$ and that of only the monopole part of $\rho_m^{\rm M}$ for $T(m)=2$ to represent the polarization of heterophilic adhesion in the two types of cells, respectively. Here, $\bm p_m$ denotes the unit vector indicating the direction of polarization in heterophilic cell-cell adhesion for the $m$th cell. In this paper, this is simply referred to as “adhesion polarity”. This polarity is an adhesion variant of that of the chemical compass during chemotaxis\cite{Bourne:2002}.
The dynamics of this concentration is assumed to be quasistatically slow and, hence, $\bm p_m$ is a slow variable. Additionally, the unit vector,  $\bm e_m (\bm r)$, represents the direction from the position of the $m$th cell, $\bm R_m$ to the peripheral position of the cell, $\bm r$, which is defined by $\bm e_m (\bm r)$ =  $(\bm r - \bm R_m)$/$|\bm r - \bm R_m|$. 
 By definition,  $\bm R_m$  is a slow coordinate variable of the $m$th cell in this expansion.  $\bm R_m$ is referred to as the center of the $m$th cell because its dynamics is assumed to be quasistatically equal to that of the center of mass of the $m$th cell. For simplicity, we assume that $\rho^{\rm M}_m$ and $\rho^{\rm D}_m$ depend only on the type function, $T(m)$. In this case, we represent $\rho^{\rm M}_m$ and $\rho^{\rm D}_m$ for $T(m) = 1$ by $\rho^{\rm M}_{T=1}$ and $\rho^{\rm D}_{T=1}$, respectively, and represent $\rho^{\rm M}_m$ for $T(m)=2$ by $\rho^{\rm M}_{T=2}$. 
The other higher order terms in the expansion are ignored because their effect is not of interest to our analysis. In this context, we have:
\begin{align}
    \rho_m(\bm r) \simeq \delta_{T(m)1}\left[\rho_{T=1}^{\rm M} + \rho_{T=1}^{\rm D}(\bm e_m(\bm r)\cdot \bm p_m)\right] + \delta_{T(m)2}\rho_{T=2}^{\rm M}.
\end{align} 

Substitution of  $\rho_m(\bm r)$ with the aforementioned expansion yields the following expression for the surface tension \cite{Matsushita:2020}:
\begin{eqnarray}
 \Gamma_{\rm Het}(\bm r, \bm r') =
 \left\{\delta_{T(m(\bm r))1}\delta_{T(m(\bm r'))2}\left[\Gamma_{np} + \Gamma_p \rho_{p}(\bm p_{m(\bm r)}, \bm r)\right] \right.\nonumber\\
  \left. +\delta_{T(m(\bm r))2}\delta_{T(m(\bm r'))1}\left[\Gamma_{np} + \Gamma_p\rho_{p}(\bm p_{m(\bm r')}, \bm r')\right]\right\}. \label{eq:interaction_hetero}
\end{eqnarray}
Here, the strength of the isotropic adhesion is $\Gamma_{np}$ = $\varepsilon(\rho^{\rm M}_{T=1}-\rho^{\rm D}_{T=1})\rho^{\rm M}_{T=2}$ and that of polarized
adhesion is $\Gamma_p$ = $\varepsilon \rho^{\rm D}_{T=1}\rho^{\rm M}_{T=2}$. 
These strengths should be restricted to induce adhesion corresponding to only positive values in $ \Gamma_{\rm Het}(\bm r, \bm r')$. 
To ensure the positivity of the strengths, the terms are redefined in this equation. For the same purpose, the positive function,  $\rho_{p}(\bm p_m, \bm r)$ = $\left[1+\bm e_{m(\bm r)}(\bm r)\cdot \bm p_{m(\bm r)}\right]$ is introduced to express the polarized component of adhesion molecule concentration and realize the tension gradient at the cell level on the tissue interface.

In this model, the adhesion polarity, $\bm p_m$, denotes the degree of freedom and its dynamics correspond to that of the adhesion molecular concentration. To analyze its dynamics, we consider the binding between the adhesion molecules in the cell membrane to the edge of the cytoskeleton related to the movement of cells. In this case, the polarization in the adhesion molecule concentration follows the direction of the movement and localizes at the leading edge of cells, as observed in experiments \cite{Coates:2001, Fujimori:2019}.
In this case, the polarity obeys the following equation \cite{Matsushita:2017}:
\begin{eqnarray}
 \frac{d \bm p_m}{d t} = \frac{1}{a \tau} \hat P(\bm p_m) \cdot \frac{d{\bm R}_m}{dt}. \label{eq:p}
\end{eqnarray}
Here, $a$ denotes the lattice constant, $t$ denotes time, and $\tau$ denotes the ratio of the relaxation time of $\bm p_m$ to that of ${\bm R}_m$. 
For the time scale, the Monte Carlo step was assumed to be the unit time. Let 
$\hat P(\bm x)$ denote the projection operator onto the direction perpendicular to a vector, $\bm x$, given by:
\begin{eqnarray}
\hat P(\bm x) = \hat 1 - \bm x^{\dagger}\otimes \bm x.
\end{eqnarray}
Here, $\hat 1$ denotes the unit tensor and $\otimes$ denotes the tensor product. This formulation is a variant of the definition of polarity for persistent walks \cite{Li:2008, Takagi:2008}.
In contrast to the latter case \cite{Szabo:2007,Fily:2012,Berthier:2014,Nagai:2015,Matsushita:2019,Matsushita:2019b}, the model cell with polarity adhesion exhibits a simple random walk in isolation owing to the absence of the driving term in the expression of free energy in this model. The motility of the cells can only be induced by heterophilic adhesive contact on interfaces of tissues following Eq.~\eqref{eq:interaction_hetero}\cite{Matsushita:2020}.

The fourth term in the RHS of Eq.~\eqref{eq:Free_enegy} is:
\begin{align}
    {\cal H}_{\rm Are} = \kappa A \sum_m \left(1 - \frac{A_m}{A}\right)^2. \label{eq:vol}
\end{align}
Here, $\kappa$ denotes the area stiffness and $A$ denotes the reference area of the area elasticity. Further, $A_m$ = $\sum_{\bm r} \delta_{m(\bm r)m}$ denotes occupation area of the $m$th cell.

Based on the existence probability,  $P(\{m(\bm r)\})$ determined on the basis of the free energy, the procedure of the proposed Monte Carlo simulation is as follows. First, the Monte Carlo simulation is used to generate a series of Potts states, $\{m(\bm r)\}$, which capture the dynamics of cell configurations. The Potts state is conventionally generated by  
16$L^2$ copies of the state, $m$, from a position, $\bm r$, to its neighboring position, $\bm r'$ \cite{Graner:1992}. 
In this case,  $\bm r$  is chosen randomly and $\bm r'$ is randomly chosen from  the nearest or  next nearest neighbor of $\bm r$. 
If a site of the scaffold is chosen, the copy is automatically rejected to maintain a fixed scaffold. Otherwise, the copy of the state is accepted by the Metropolis probability, $\min \{1, P(\{m(\bm r')\})/P(\{m(\bm r)\}) \}$. Here, 
$P(\{m(\bm r)\})$ and $P(\{m(\bm r')\})$  denote the Boltzmann factor of the state preceding the copy and that of a candidate of the update state following the copy, respectively. 
This set of copies is called a single Monte Carlo step (MCs). Following this single Monte Carlo step, the adhesion polarity is updated once by integrating Eq.~\eqref{eq:p} via the Euler method. Simultaneously, the center of the cell, $\bm R_m$ is also updated using  $\bm R_m$ = $\sum_{\bm r}\delta_{m(\bm r)m} \bm r/A_m$.

\section{Simulation and Results}

\begin{figure}[t]
    \begin{center}
        \includegraphics[width=1.0\linewidth]{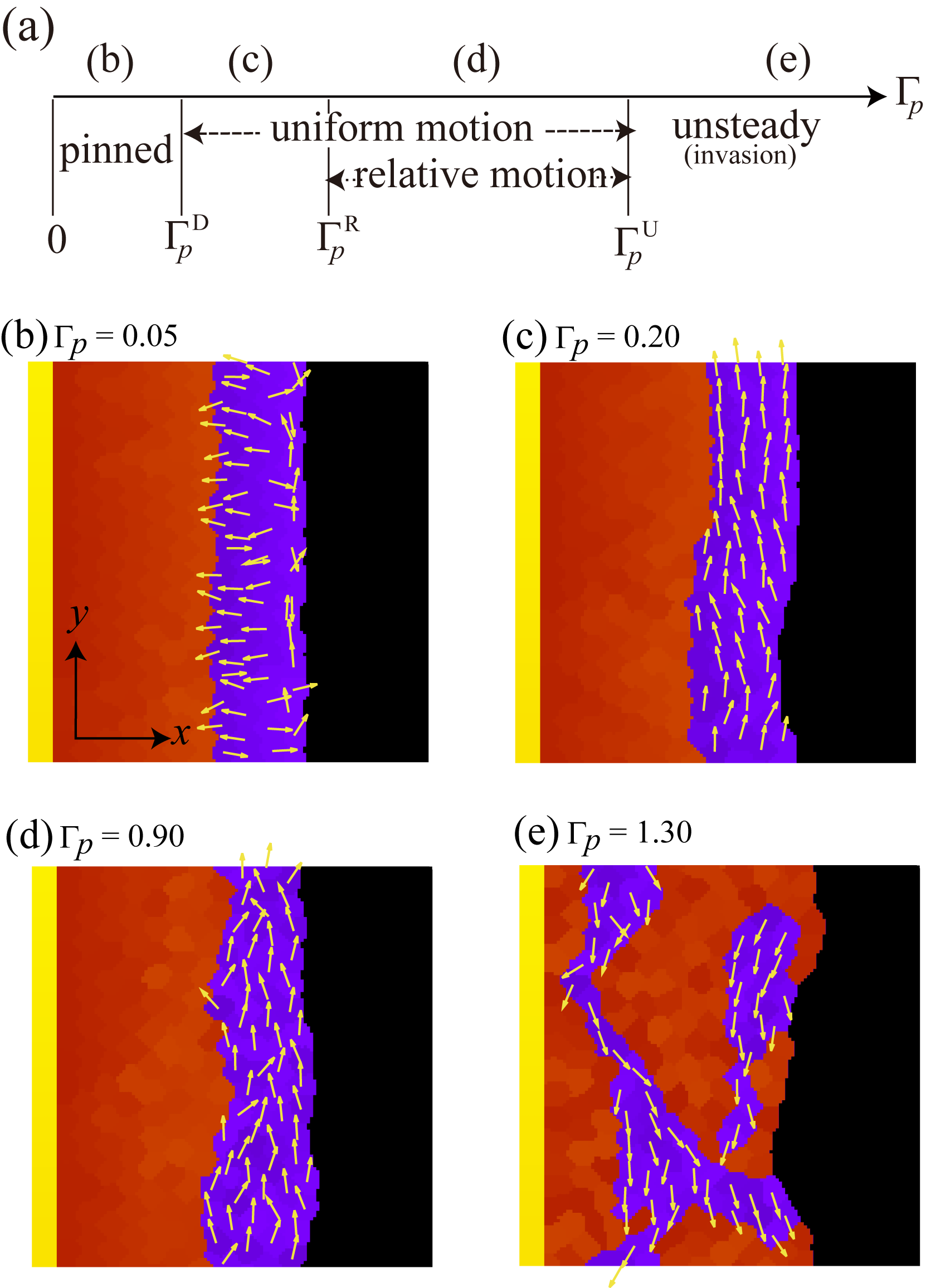}
        \caption{(Color online) (a) Schematic phase diagram. (b-e) Configuration snapshots of cells and polarities for (b) the pinned state corresponding to $\Gamma_p$ = 0.05, (c) the uniformly moving state corresponding to  $\Gamma_p$ = 0.20, (d) the relatively moving state corresponding to  $\Gamma_p$ = 0.90, and (e) the unsteady state corresponding to $\Gamma_p$ = 1.30.  The yellow, red, and violet regions represent the tissues comprising the scaffold, cells with  $T(m)$ = 2, and cells with $T(m)$ = 1, respectively. The black region denotes empty space. Arrows on the cells represent the adhesion polarity. }
        \label{fig:states}
    \end{center}
\end{figure}

In this section, we examine the emergence of Marangoni flow on the one-dimensional tissue interface under predefined conditions. In the first half, the simulation configuration for this analysis are explained in detail. The simulation size is determined as follows. With regard to the space of the system, we assume the lattice to be a two-dimensional square lattice for simplicity and the $x$- and $y$-directions axes are assumed to be the lattice axes.
The linear system size is taken to be $L$=128$a$ equipped with a periodic boundary condition to realize tractable simulation. 
 The cell flow on the interface is expected in the case of flat tissue, because of the possible adhesion of the rough interface to the tissues. To realize an almost flat interface between the two tissues, the area of each cell is taken to be $A$ = 64$a^2$  combined with the number of cells:
 The number of cells with $T(m)$ = 1 is taken to be  $N_1$ = 64 and the number of those with $T(m)$ = 2  is taken to be $N_2$ = 112.
 In particular, this configuration realizes a flat interface along the axes even corresponding to heterophilic adhesion of low strength (See Fig.~\ref{fig:states}(b)).

To realize stable model cells, the following parameters are chosen. To achieve polarity dynamics of  $T(m)$ = 1, a large value of  $\tau$ = 5.0 is selected, which reproduces cell movements  \cite{Matsushita:2020}.
The area stiffness of $\kappa$ in Eq.~\eqref{eq:vol} is taken to be a large value, 10, to ensure the stable maintenance of the cellular area. To easily obtain adhesion propulsion of cells alongside stability of cellular area, $\beta$ should be selected within an optimal range. In this simulation, we empirically set $\beta$ = 0.5 based on a previous work \cite{Matsushita:2020}. 
This value is also suitable for maintaining the flat interface corresponding to the following tension parameters.

\begin{figure}[t]
    \begin{center}
        \includegraphics[width=1.0\linewidth]{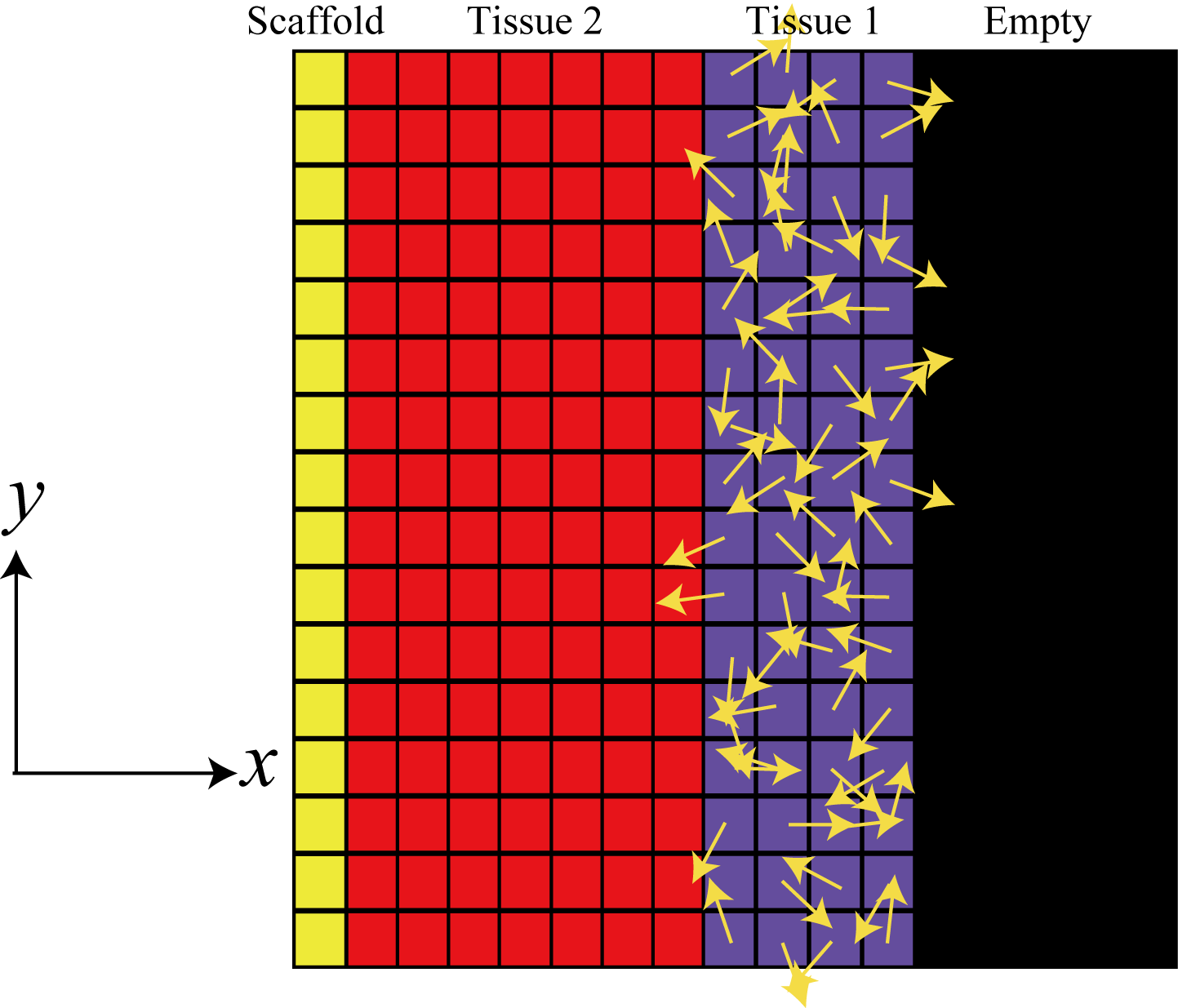}
        \caption{(Color online) The initial Potts state with adhesion polarities used in our simulation. The yellow, red, and violet domains represent the initial configuration of the scaffold, cells with $T(m)$ = 2, and cells with $T(m)$ = 1, respectively. The black region denotes empty space. Arrows on the cells represent the adhesion polarities associated with individual cells with  $T(m)$ = 1.}
        \label{fig:initial}
    \end{center}
\end{figure}
\begin{figure*}[t]
    \begin{center}
        \includegraphics[width=1.0\linewidth]{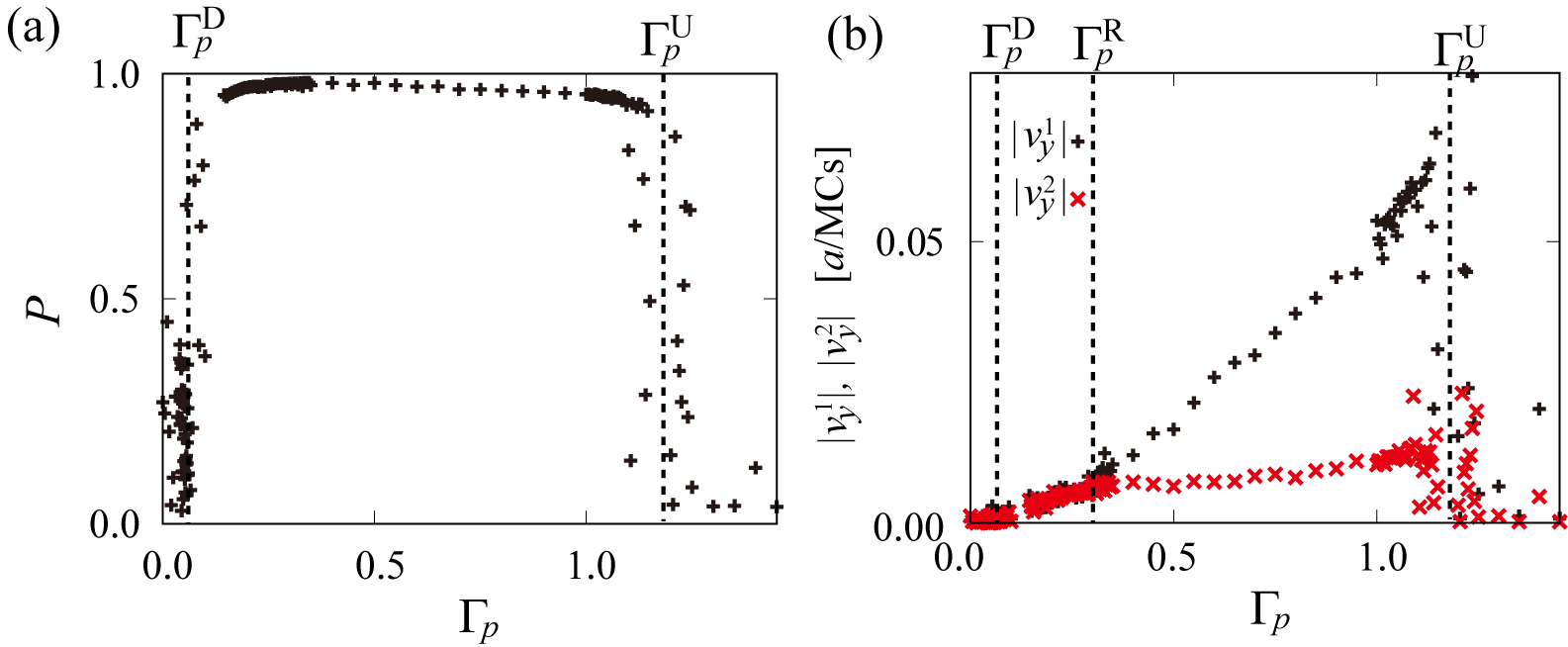}
        \caption{(Color online) (a)  The order parameter  $P$ as a function of $\Gamma_p$. (b) The average collective velocity for $T(m)=1$ ($+$), $v_y^1$, and that for $T(m)=2$ ($\times$), $v_y^2$.}
        \label{fig:order_parameter}
    \end{center}
\end{figure*}
 
Now, let us consider the tension parameters in free energy. These parameters are estimated to stabilize the layered tissue structure with a flat interface, as depicted in Fig.~\ref{fig:initial}.
To consider the contact states of cells with empty spaces and scaffolds, we set the values of tensions as follows.  $\Gamma_{\rm E}$ = 6.0.  is taken to be the base value of tension. Therefore, the cells with 
$T(m)$ = 1 and 2 form tissues when the tensions between pairs are smaller than 2$\Gamma_{\rm E}$ = 12.0.
We set both $\Gamma^{1}$ = 4.0 and $\Gamma^{2}$ = 4.0 are taken to stabilize the respective tissues. In addition, because intermixing of tissues destabilizes them. Further, $\Gamma_{\rm I}$ is required to be larger than half of $\Gamma^1$ and $\Gamma^2$ because intermixing of tissues destabilizes them.
Further, $\Gamma_{\rm I}$  is also required to be smaller than $\Gamma_E$ to stabilize the interface between the tissues against invasions of empty spaces between them. To satisfy the aforementioned stability conditions,  $\Gamma_{\rm I}$ = 4.0 is taken. 
 To eliminate the effect of the nonpolarized part of heterophilic adhesion, $\Gamma_{np}$ = 0.0 
 is taken. Various values of  $\Gamma_p$ are used to investigate the permissible range of adhesion that promotes cell flow between tissues. As mentioned previously, cells with $T(m)$ = 1 are assumed to be incapable of forming adhesive contact with scaffolds and those with $T(m)$ = 2 are assumed to form adhesive contact with scaffolds. To reproduce this situation, we take $\Gamma_{\rm S}^1$ = 13.0 and $\Gamma_{\rm S}^2$ = 4.0.

The initial state of the simulation is schematically depicted in Fig.~\ref{fig:initial}.
In this state,  16 cells are aligned along the $y$-direction at
the left edge of the system (around  $x$ = 1) ), constituting a scaffold. The array spans the range from $y$ = 1 to $y$ = $L$ in the $y$-direction. 
The scaffold cells do not move and thereby inhibit any cell movement passing through themselves. On the right side of the array, cells with $T(m)$ = 2 form 7 $\times$ 16  array and constitute a tissue with a band structure. The tissue connects with itself between $y$ = 1 and $y$ = $L$  in the periodic boundary condition. The left sides of these arrays are adhered to the array of scaffold cells. On the right side of this tissue, cells with $T(m)$ = 1 form 4 $\times$ 16 array and constitute a tissue with a band structure similar to that of cells with $T(m)$ = 2. 
 In these cells, the initial directions of adhesion polarities are random. From this initial state, a steady state is attained via simulation over $t_0$ = $10^5$ MCs and the dependence of the steady states on $\Gamma_p$  is thereby analyzed.

To examine the emergence of the Marangoni flow, we calculate an observable metric under the aforementioned configuration. The observable metric relevant to probing the emergence of the Marangoni flow is an order parameter of the adhesion polarity, $\bm p_m$, defined as follows:
\begin{eqnarray}
P = \left| \frac{1}{N_1} \int_{t_0}^{t_1+t_0}dt \sum_{m \in \Omega_1} \bm p_m(t) \right|,
\end{eqnarray}
where $\Omega_1$  denotes the set of indices for cells with $T(m)$ = 1 and $t_1$denotes the Monte Carlo step used to average the order parameter. We take 10$^6$ MCs as $t_1$.
When the value of the order parameter becomes almost unity, the emergence of polarity order is confirmed. The order is a necessary condition for the existence of adhesion-induced collective cell flow \cite{Matsushita:2018}. In turn, the Marangoni flow can be investigated based on this collective flow.

We calculate the values of the aforementioned parameter with respect to varying values of $\Gamma_p$ to identify the emergence of the Marangoni flow.
 If heterophilic adhesion drives the flow around the tissue interface, large values of $P$  are expected to correspond to large values of $\Gamma_p$ \cite{Kabla:2012,Matsushita:2020}.
To verify this, $P$ is plotted in Fig.~\ref{fig:order_parameter}(a) as a function of $\Gamma_p$.
It is evident that the order parameter takes significantly small values when $\Gamma_p^{\rm D}$ $\sim$ 0.1 or lower. When  $\Gamma_p$ exceeds $\Gamma_p^{\rm D}$, $P$ becomes almost unity. Therefore, heterophilic adhesion of high strength at least stabilizes the order of adhesion polarity and may drive the collective motion of the cells with $T(m)$ = 1. 
When $\Gamma_p$ exceeds $\Gamma_p^{\rm U}$ $\sim$ 1.2, the value of $P$ decreases again, indicating the vanishing of the order. 

When $\Gamma_p$  lies within $\Gamma_p^{\rm D}$ and $\Gamma_p^{\rm U}$, the collective motion of cells with $T(m)$ = 1 is expected. In this range, the collective motion can be of two types--uniform motion over two tissues and relative motion between two tissues. Therefore, large values of $P$ are not directly correlated to relative motion between two tissues as in the case of the Marangoni flow. To directly verify relative motion, the velocities of the two tissues is required to be monitored.

To this end, we calculate the average velocities of both tissues and plot them in Fig.~\ref{fig:order_parameter}(b). Here, the average velocity corresponding to each tissue is aligned along the $y$-direction because of the geometry of the tissue, and so, only the $y$-component of velocities are plotted. The average velocity of a tissue with  $T(m)$ = $T$ is defined as follows:
\begin{align}
    v^T_y = \frac{1}{N_T}\sum_{m \in \Omega_{T}} \int_{t_0}^{t_1+t_0} dt d_y^m.
\end{align}
Here $d_y^m$ denotes the displacement of the $m$th cell per MCs in the $y$-direction and $\Omega_T$ denotes the set of indices for cells with $T(m)$ = $T$. 
The two average velocities, $v_y^1$ and $v_y^2$, in the range from $\Gamma_p^{\rm U}$ to $\Gamma_p^{R}$ $\sim$ 0.3 are almost identical.
This corresponds to uniform motion over the two tissues in this range of $\Gamma_p$.
In contrast, in the range from $\Gamma_p^{\rm R}$ to $\Gamma_p^{\rm U}$, the average velocities of the two tissues are not only finite but also distinct. This indicates relative motion between the two tissues and, consequently, the Marangoni flow. In particular, $v_y^1$ increases with increasing $\Gamma_p$ unlike $v_y^2$, which remains more or less constant in this range of  $\Gamma_p$.
This indicates that the enhancement of heterophilic adhesion accelerates the flow.

\begin{figure}[t]
    \begin{center}
        \includegraphics[width=1.0\linewidth]{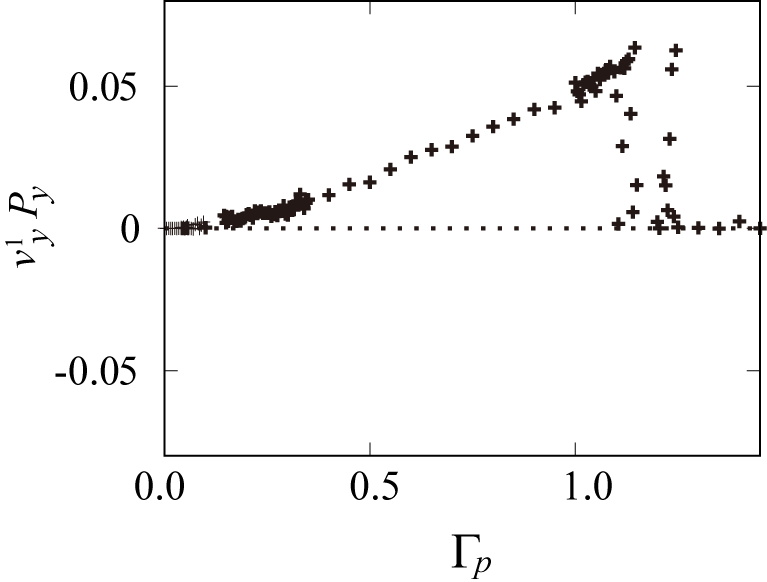}
        \caption{The product $P_yv_y^1$ as  a function of $\Gamma_p$.}
        \label{fig:direction}
    \end{center}
\end{figure}

Next, we focus on the direction of the observed flow. The Marangoni flow is oriented in the direction opposite to that of the tension gradient, i.~e., from low to high tension  \cite{Getling:1998}.
 If the observed flow shares its origin with the Marangoni flow, the direction of this flow would also be opposite to that of the tension gradient. To estimate its direction, the adhesion polarity, $\bm p_m$ having the same direction as the tension gradient can be used. In particular, by using $\bm p(\bm x)$ = $\sum_m \bm p_m \delta(\bm R_m - \bm x)$ with the delta function $\delta(\bm x)$,
\begin{align}
    - \nabla \gamma(\bm x) \propto \sum_m \bm p_m \delta(\bm R_m - \bm x)
\end{align}
can be naively expected from Eq.~\eqref{eq:polarization-tension}.
In this case, the $y$-component of the average $\bm p_m$,
\begin{eqnarray}
P_y = \frac{1}{N_1} \sum_{m \in \Omega_1}\int_{t_0}^{t_1+t_0} dt  p_m^y
\end{eqnarray}
and $v^1_y$ are expected to have opposite signs by Eq.~\eqref{eq:Marangoni_flow}. Hence, their product is expected to be negative by naive speculation. Here, $p_m^y$ denotes the $y$-component of $\bm p_m$.
To verify the negativity of the product,  $P_yv^1_y$ is plotted in Fig.~\ref{fig:direction}.
Unexpectedly, the product is observed to be positive, which indicates that the tension gradient drives the cell movement along its own direction. Thus, the tissue interface tension, $\gamma(\bm x)$, seemingly induces the cell flow velocity, $\bm v_{\rm Cell}(\bm x)$, given by
\begin{eqnarray}
\bm v_{\rm Cell}(\bm x) \propto - \nabla \gamma(\bm x),
\end{eqnarray}
which differs from that in Eq.~\eqref{eq:Marangoni_flow}.
This prominent difference between the observed flow and the Marangoni flow indicates the existence of a different microscopic mechanism in the effect of the tension gradient.

Now, we summarize the states identified thus far using the phase diagram depicted in Fig.~\ref{fig:states}(a). The first state corresponds to when $\Gamma_p$ is lower than $\Gamma_p^{\rm D}$ which induces no collective motion and cells are pinned. Here, $\Gamma_p^{\rm D} $ denotes $\Gamma_p$ for the depinning of collective motion. The second state corresponds to when $\Gamma_p$ lies between  $\Gamma_p^{\rm D}$ and $\Gamma_p^{\rm R}$, which induces uniform motion over the two tissues.  Here,  $\Gamma_p^{R}$ denotes the threshold of $\Gamma_p$ for the inducement of relative motion. The third state corresponds to when $\Gamma_p$  lies between $\Gamma_p^{\rm R}$ and $\Gamma_p^{\rm U}$, which induces relative motion between the two tissues.  Here, $\Gamma_p^{\rm U}$ acts as an unstable point for collective motion. When $\Gamma_p$  exceeds $\Gamma_p^{\rm U}$, the state transitions to one without motion ordering. In order,  the aforementioned states are referred to as a pinned state, a uniformly  moving state, a relatively moving state, and an unsteady state. Of these, the characteristics of the unsteady state are not completely known yet.

To clarify the characteristics of the unsteady state, we calculate the cell configurations and polarities in the four typical states, as illustrated in Figs.~\ref{fig:states}(b)-\ref{fig:states}(e). 
The pinned state corresponds to $\Gamma_p$ = 0.05, as depicted in  Fig.~\ref{fig:states}(b); the uniformly moving state corresponds to $\Gamma_p$ = 0.20 as recorded in Fig.~\ref{fig:states}(c); and the relatively moving state corresponds to $\Gamma_p$ = 0.90, as shown in Fig.~\ref{fig:states}(d). Further, all three exhibit the same layered tissue structure depicted in Fig.~\ref{fig:initial} in the initial state. These observations indicate that the layered structure of tissues is stable against the driving force of collective cell movement. In particular, the unsteady state corresponding to $\Gamma_p$ = 1.30, as illustrated in  Fig~\ref{fig:states}(e), exhibits a mixing of tissues  1 and 2. This indicates that strong heterophilic adhesion destabilizes the two tissues by overcoming the free energy barrier required to roughen the interface between them. Further, the domain shapes of tissues are complex in the unsteady state, which introduces a high degree of randomness into the collective motion. This reflects in significant fluctuations in $v_y^1$ and $v_y^2$  in Fig.~\ref{fig:order_parameter}(b) when $\Gamma_p$ exceeds $\Gamma_p^{\rm U}$.

\section{Summary and Discussion}
In this paper, we examined 
the possibility of “cell Marangoni flows” in the cell-scale surface tension gradient. As per expectations, we confirmed relative motion between two tissues induced by the tension gradient as in the case of the Marangoni flow. However, the direction of this flow was observed to be in the opposite direction to that of the Marangoni flow. Further, the flow state was only observed within an optimal range of strength of heterophilic adhesion, which was ascertained to be determined by the thresholds, $\Gamma_p^R$ and $\Gamma_p^{\rm U}$.

Now, we discuss two characteristics of this flow. The first concerns the emergence conditions of the relatively moving state with respect to heterophilic adhesion. As its emergence is determined by the threshold values,  $\Gamma_p^{\rm R}$ and $\Gamma_p^{\rm U}$, they are now discussed further. First, let us consider $\Gamma_p^{\rm R}$.
Figure~\ref{fig:order_parameter}(c) depicts the acceleration of  $v_y^2$ with respect to $v_y^1$ in the range between $\Gamma_p^{\rm D}$ and  $\Gamma_p^{\rm R}$ in contrast with that in the range between $\Gamma_p^{\rm R}$ and $\Gamma_p^{\rm U}$.
This indicates that by forming stable solid-like arrangements on the interface, cells in the first tissue drug the cells in the second tissue, which induces uniform motion in the uniformly moving state. Further, as illustrated in Fig.~\ref{fig:states}(d), the interface becomes rougher than that shown in Fig.~\ref{fig:states}(c). These observations imply a phase change to liquid surrounding the interface. Based on this information, $\Gamma_p^{\rm R}$  is expected to pin down the interface, which may result from the stability of the solid-like alignment of cells surrounding the interface in the uniformly moving state. Therefore,  $\Gamma_p^{\rm R}$ corresponds to the induced melting of cell alignments on the interface. However, the theoretical evaluation of $\Gamma_p^{\rm R}$ remains elusive. 

\begin{figure}[t]
    \begin{center}
        \includegraphics[width=1.0\linewidth]{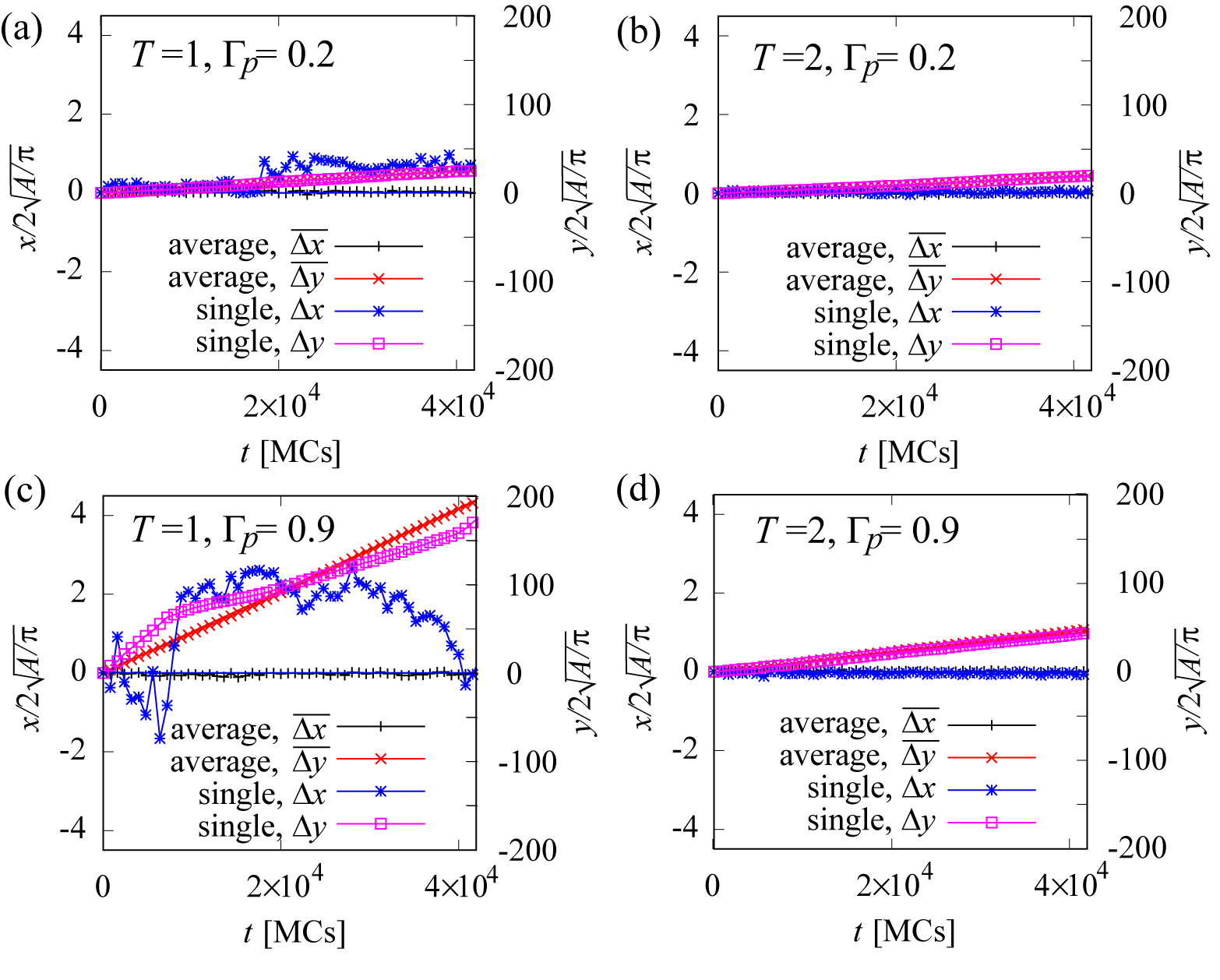}
        \caption{(Color online) $x(t)$ and $y(t)$ for a uniformly moving state with (a) $T=1$ and $\Gamma_p$=0.2; (b)  $T=2$ and $\Gamma_p$=0.2, and for a relatively moving state with (c) $T=1$ and $\Gamma_p$=0.9; (d) $T=1$ and $\Gamma_p$=0.9. The origin ($t=t_0$) is set zero for all x(t) and y(t). The symbols $+$, $\times$, $+$ \hspace{-3.0mm} {$\times$}, $\boxdot$  denote the average $x$ over all cells in the tissue, the average  $y$  coordinate over all cells in the tissue, the $x$ coordinate of a single cell, and the  $y$ coordinate of a single cell, respectively.}
        \label{fig:displacements}
    \end{center}
\end{figure}

This melting can also be indirectly confirmed by the difference between the average movement of cells and that of a single cell in the same tissue in the uniformly and relatively moving states. In  Figs.~\ref{fig:displacements}(a)-\ref{fig:displacements}(d), the displacements of a single cell, $\Delta x(t)$ and $\Delta y(t)$, are plotted with those of average cells, $\overline{ \Delta x(t)}$ and $\overline{\Delta y(t)}$. Here, their origin are set to zero at $t$ = 0, and they are a function of $t$ during a short period when $t$ = 0 corresponds to $t = t_0$ in simulation.
In the uniformly moving state corresponding to values below $\Gamma_p^{\rm R}$, $\Gamma_p =0.20$ is chosen.  The positions of the average and single cells are observed to exhibit identical behavior for $T$=1, as illustrated in Fig.~\ref{fig:displacements}(a), and $T=2$, as illustrated in  Fig.~\ref{fig:displacements}(b), except for $\Delta x(t)$ in $T$ = 1. Even $\Delta x(t)$ of a single cell in $T$ = 1, it at most is confined in a range of a single cell size displacement from that of the average cells, namely $|\Delta x(t) - \overline{\Delta x(t)}| \lesssim 2\sqrt{A/\pi}$. These observations indicate that the cells behave as a uniform solid in this case. Unlike the uniformly moving state, in the relatively moving state, the $T=1$ tissue is fluidized.  To verify this, the motion of $T=1$ is plotted in Fig.~\ref{fig:displacements}(c)  and that of $T=2$ is plotted in  Fig.~\ref{fig:displacements}(d) during the relatively moving state corresponding to $\Gamma_p$ = 0.9. 
As depicted in Fig.~\ref{fig:displacements}(c), a single cell actively moves in the $x$-direction beyond the single cell size and accordingly varies its velocity in the $y$-direction, as evidenced by the slope of $\Delta y(t)$.  This behavior is prominently different from those of average cells. This difference implies that the tissue $T=1$ is in liquid phase and its velocities depend on the position in the $x$-direction. In particular, the position $\Delta x(t)$ is observed to be negatively correlated to the slope of $\Delta y(t)$. Therefore, owing to the melting of the tissue $T=1$, a laminar flow is expected, which exhibits high velocity near the interface. Further, single cell remain stable corresponding to positions with high $\Delta x(t)$ compared to those with low $\Delta x$ positions. Because the values of $\Delta x(t)$ decreases with increasing proximity of the cell to the interface, this indicates melting at positions near the interface.

Now, let us consider $\Gamma_p^{\rm D}$. It is expected to be the threshold for the invasion of cells with $T(m)$ = 1 to $T(m)$ = 2 because of the mixing of tissues illustrated in Fig.~\ref{fig:states}(e).
The threshold for the invasion is estimated from $\Gamma_p^{\rm U}$ $\gtrsim$ $[\Gamma_{\rm I} - \Gamma_1/2]$ / $\max_{\bm r}\rho_p(\bm p_m, \bm r)$ $\simeq$ 1.0 via underestimation using the maximum strength for heterophilic cell-cell adhesion.
This estimated value is consistent with the observed value of 1.2 depicted in Fig.~\ref{fig:order_parameter}(b) and directly confirms that $\Gamma_p^{\rm U}$ is determined by the invasion.

The other characteristic to be discussed is the direction of the flow, which is opposite of that of the Marangoni flow. This direction is the only natural feature commonly associated with a self-propelled droplet \cite{Levan:1981,Wasan:2001}, and is induced by cell movement in cell-scale surface tension. The most important feature in the mechanism of cell movement that contributes to the difference is the fact that cells use shape deformations during the movement. The peripheral part of the cell, i.e., that corresponding to low tension and high adhesion, is easily extensible \cite{DiMilla:1991,Huttenlocher:1995, Matsushita:2018, Okuda:2021}.
Therefore, the cell moves from regions with high tension to those with low tension. This is in contrast with the direction of the Marangoni flow, which is from low to high tension. This explains the difference between the flow directions.

Lastly, we consider the possibility of confirming the discussed scenario using experimental observations. The conclusions of this paper are based on the existence of polarization in the concentration of heterophilic adhesion molecules. Thus, the hypothesis may be verified by directly examining this polarization. In addition, this paper predicts the existence of a threshold for the strength of adhesion that determines the existence of relative motion between tissues. This prediction may be effectively confirmed by observing slug velocity in a light strength control \cite{Poff:1973,Poff:1984}. If phototaxis is based on the polarization of heterophilic adhesion, the prediction stated in this paper indicates the existence of a threshold of light strength for the existence of phototaxis in dicty.

For the experimental confirmation, clarification of the molecular basis for heterophilic adhesion is important to support the experimental idea. A possible candidate for the heterophilic adhesion molecules are TgrB1 and TgrC1, which is manifested in dicty slug  \cite{Siu2004}. Further, the adhesion distribution on the cell membrane exhibits a polarization \cite{Fujimori:2019}. This polarization of these adhesion molecules is the preferred feature for inducing cell Marangoni effect on tissue interfaces. However, the discussed scenario assumes the manifestation of different heterophilic adhesion molecules in the two tissues. In the case of dicty, this corresponds to the situation in which TgrB1(or C1) acts only in the prestalk region of the slug and TgrC1(or B1) acts only in the remaining region (See Fig.~\ref{fig:phototaxis}(a)).  
 To the best of our knowledge, such separated region-specific action of heterophilic adhesion molecules has not been observed. At the very least, the genes encoding TgrB1 and TgrC1 have a common promoter region and are usually transcribed simultaneously  \cite{Hirose:2017}. 
Therefore, the direct confirmation of cell Marangoni flow should need the direct confirmation of the insitu functional adhesion activity of TgrB1 and TgrC1 on cell membrane.

\begin{acknowledgments}
We thank the support on the research resource by M. Kikuchi and H. Yoshino. We also thank I. Shibano for helpful comments on the coexpression of TgrB1/TgrC1 and thank S. Yabunaka for information of self-propelled droplets with an interface tension. This work is supported by JSPS KAKENHI (Grant Number 19K03770, 18K13516) and by AMED (Grant Number JP19gm1210007).
We acknowledge utilizing the supercomputers of ISSP.
\end{acknowledgments}
%

\end{document}